\documentclass{desyproc}

\begin{document}
\title{Supernova neutrino flavor evolution at high densities}

\author{{\slshape A.B. Balantekin$^{1,2}$}\\[1ex]
$^1$University of Wisconsin, Department of Physics, Madison, WI  53706 USA\footnote{Permanent address.}\\[1ex]
$^2$Nuclear Science Division, Lawrence Berkeley National Laboratory, Berkeley, CA 94720 USA}

\contribID{xy}

\confID{1964}  
\desyproc{DESY-PROC-2010-01}
\acronym{HANSE2011} 
\doi  

\maketitle

\begin{abstract}
These conference proceedings cover various aspects of neutrino propagation through the high 
matter and neutrino densities near the proto-neutron star in a core-collapse supernova. A significant feature of this regime is the impact of neutrino-neutrino interactions.  Properties of this non-linear many-neutrino system are discussed with a particular emphasis on its symmetries. 
\end{abstract}

\section{Introduction}
Almost all the gravitational binding energy of the pre-supernova star is converted into neutrinos and antineutrinos in a core-collapse supernova, yielding a very large neutrino flux \cite{Fuller:1998kb}. Consequently neutrino properties play a very important role, not only in the dynamics of a core-collapse supernova, but also in the r-process nucleosynthesis if supernovae are the appropriate sites \cite{Balantekin:2003ip}. Neutrinos traveling through supernovae undergo matter-enhanced neutrino oscillations (due to the MSW effect resulting from neutrinos interacting with the background electrons and positrons) much like neutrinos traveling through the Sun or the Earth \cite{fullerold}. However, unlike these latter sites, in a supernova environment it is possible to have  matter-enhanced antineutrino flavor transformations \cite{Qian:1995ua}. It was ascertained that, because of the large number of neutrinos ($\sim 10^{58}$) emitted by the proto-neutron star, background neutrinos also contribute to the coherent forward scattering of neutrinos in a core-collapse supernova 
\cite{fullerold,Notzold:1987ik}. Once the importance of the flavor-mixing non-diagonal terms coming from neutrino-neutrino interactions was highlighted \cite{Pantaleone:1992eq}, it became clear that one deals with a genuine many-body problem with one- and two-body interactions. 

The significance of the neutrino-neutrino interactions in core-collapse supernovae  \cite{Qian:1994wh} and the possibility of the occurrence of collective effects due to those interactions were recognized early on \cite{Pastor:2002we}. 
In a supernova environment neutrino-neutrino 
interactions \cite{Pantaleone:1992eq}
play a crucial role both for neutrinos and antineutrinos 
\cite{Qian:1994wh,Pastor:2002we,Friedland:2003dv,Gava:2009pj}. 
Since such collective neutrino oscillations dominate the neutrino propagation much deeper than the conventional matter-induced MSW effect, it would impact the r-process nucleosynthesis 
\cite{Balantekin:2004ug,Chakraborty:2009ej,Duan:2010af}. There is an extensive literature on this subject \cite{sample}, a good starting point is several recent surveys 
\cite{Duan:2009cd,Duan:2010bg,Raffelt:2010zz}. 
 An algebraic approach to this problem was worked out in Ref. \cite{Balantekin:2006tg} from a many-body point of view. Such an algebraic approach is helpful in exploring the hidden symmetries of the system. Hamiltonian describing collective neutrino oscillations possesses an $SU(N)_f$ rotation symmetry in the neutrino flavor space \cite{Balantekin:2006tg,Duan:2008fd,Balantekin:2009dy}. Various collective modes, including spectral swappings or splittings arise from this symmetry even in the inhomogeneous  or anisotropic environments \cite{Duan:2008fd}. One expects that such a complex nonlinear system may exhibit further symmetries. Indeed, several authors noted the presence of various conserved quantities in the collective neutrino oscillations \cite{Raffelt:2007cb,Duan:2007mv}.  
More recently, it was shown that collective oscillations that maintain coherence can be classified by a number of linearly-independent functions \cite{Raffelt:2011yb}, implying that scalar products of a unique linear combination of the original polarization vectors  are conserved. The flavor evolution of a dense neutrino system by taking into account both the vacuum oscillations and self interactions of neutrinos from a many-body perspective was considered in Ref. \cite{Pehlivan:2010zz}.  
Using the similarity between the collective neutrino oscillation Hamiltonian and the BCS Hamiltonian 
one can show that, in the single angle approximation, both the full many-body picture and the commonly-used effective one-particle picture possess several constants of motion \cite{Pehlivan:2011hp}. 

One appealing aspect of the core-collapse supernovae is that they are the laboratories where diverse aspects of neutrino physics come into play.  Here we concentrate on the collective behavior arising from the neutrino-neutrino interactions in a supernova and omit other interesting topics such as the role of sterile neutrinos \cite{McLaughlin:1999pd} or the neutrino magnetic moment \cite{Balantekin:2007xq} in the 
r-process nucleosynthesis, effects of the CP-violation \cite{Balantekin:2007es}, effects of turbulence and density fluctuations \cite{Loreti:1995ae}, the role of neutrinos in shock revival \cite{Dasgupta:2011jf}, and neutrino signatures of black hole formation \cite{Sasaqui:2005rh}. 

The next chapter of this proceedings contribution describes the algebraic formulation of the neutrino-neutrino interactions in the many-neutrino system. Chapter 3 includes a discussion of the invariants of the Hamiltonian of this system. Brief concluding remarks are included in Chapter 4. 

\section{An Algebraic Formulation of the Neutrino-Neutrino Interactions}

For simplicity, we first consider a neutrino gas with two flavors and no antineutrinos. Matter and flavor basis creation and annihilation operators for a neutrino with momentum $\mathbf{p}$ and spin $s$ are related as 
\begin{equation}
\label{1}
a_{1}(\mathbf{p},s)  =  \cos\theta\: a_{e}(\mathbf{p},s)-\sin\theta\: a_{x}(\mathbf{p},s)
\end{equation}
\begin{equation}
\label{2}
a_{2}(\mathbf{p},s) =  \sin\theta\: a_{e}(\mathbf{p},s)+\cos\theta\: a_{x}(\mathbf{p},s) .
\end{equation}
It is easy to show that the flavor isospin operators defined as 
\[
\hat{J}_{\mathbf{p},s}^{+}= a_{e}^{\dagger}(\mathbf{p},s)a_{x}(\mathbf{p},s)~,\qquad
\hat{J}_{\mathbf{p},s}^{-}= a_{x}^{\dagger}(\mathbf{p},s)a_{e}(\mathbf{p},s)~,\qquad
\]
\[
\hat{J}_{\mathbf{p},s}^{0}=\frac{1}{2}\left(a_{e}^{\dagger}(\mathbf{p},s)a_{e}(\mathbf{p},s)-a_{x}^{\dagger}(\mathbf{p},s)a_{x}(\mathbf{p},s)\right) 
\]
form an SU(2) algebra:
\[
[\hat{J}_{\mathbf{p},s}^{+},\hat{J}_{\mathbf{q},r}^{-}]=2\delta_{\mathbf{p}\mathbf{q}}\delta_{sr}\hat{J}_{\mathbf{p},s}^{0}~,\qquad
[\hat{J}_{\mathbf{p},s}^{0},\hat{J}_{\mathbf{q},r}^{\pm}]=\pm\delta_{\mathbf{p}\mathbf{q}}\delta_{sr}\hat{J}_{\mathbf{p},s}^{\pm}~.
\]
One can show that the particle mixing given in Eqs. (\ref{1}) and (\ref{2}) is generated by this algebra:
\[
\hat{U}^\dagger a_1(\mathbf{p},s)\hat{U}  = a_e(\mathbf{p},s) 
\]
with 
\begin{equation}
\label{mixalg}
\hat{U} =e^{\sum_p \tan \theta J_p^+} \; e^{- \sum_p \ln(\cos^2 \theta) J_0^p} \; e^{-\sum_p \tan \theta J_p^-}  .
\end{equation}

After subtracting a term proportional to the identity, the one-body Hamiltonian including interactions with the electron background takes the form
\[
\hat{H}_\nu  = \sum_p \left( \frac{\delta m^2}{2p} \hat{B}\cdot\vec{J}_p
 - \sqrt{2} G_F 
N_e  J_p^0  \right)
\]
where one defines 
\[
\hat{B} = (\sin2\theta,0,-\cos2\theta).
\]
The neutrino-neutrino interaction term in the Hamiltonian is 
\begin{equation}
\label{nunu}
\hat{H}_{\nu \nu} 
= 
\frac{\sqrt{2}G_{F}}{V}\sum_{\mathbf{p},\mathbf{q}}\left(1-\cos\vartheta_{\mathbf{p}\mathbf{q}}\right)\vec{J}_{\mathbf{p}}\cdot\vec{J}_{\mathbf{q}}
\end{equation}
where $\vartheta_{\mathbf{p}\mathbf{q}}$ is the angle between neutrino momenta 
$\mathbf{p}$ and $\mathbf{q}$. In Eq. (\ref{nunu}) $(1 - cos \vartheta)$ terms follow from the V-A nature of the weak interactions.

The evolution operator 
\[
i\frac{\partial {\cal U}}{\partial t} = \left( H_{\nu} + H_{\nu \nu} \right)  {\cal U} 
\]
can be calculated \cite{Balantekin:2006tg} as a path integral using SU(2) coherent states:  
\[
|z(t)\rangle = \exp{\left(\int dp z(p,t) J_+(p) \right)} |\phi 
\rangle , \>\>\> 
|\phi \rangle =\prod_p a_e^\dagger(p)|0 \rangle \nonumber
\]
to obtain
\begin{equation}
\langle z'(t_f)| {\cal U} |z(t_i) \rangle = \int {\cal D}[z,z^*] 
\exp{\left(iS[z,z^*]\right)} \nonumber
\end{equation}
where 
\begin{equation}
S(z,z^*)= \int_{t_i}^{t_f}dt \frac{ \langle z(t)|i\frac{\partial}{ 
\partial t}-H_{\nu} - H_{\nu \nu})|z(t) \rangle} {\langle z(t)|z(t) \rangle } + \log 
\langle z'(t_f)|z(t_f) \rangle \nonumber .
\end{equation}
The stationary phase approximation to the path integral 
\begin{equation}
\left(\frac{d}{dt}\frac{\partial}{\partial \dot{z}}-
\frac{\partial}{\partial z}\right)L(z,z^*)=0 \ \ \ \ \ \
\left(\frac{d}{dt}\frac{\partial}{\partial \dot{z}^*}-
\frac{\partial}{\partial z^*}\right)L(z,z^*)=0 \nonumber
\end{equation}
yields the differential equation 
\begin{equation}
i\dot{z}(p,t)=\beta(p,t)-\alpha(p,t) z(p,t)-\beta^*(p,t) z(p,t)^2 
\label{zeq}
\end{equation}
where we defined 
\begin{equation}
\alpha(p,t) = -\frac{\delta m^2}{2p}\cos{2\theta}+\sqrt{2} G_F
N_e + \sqrt{2}G_F \int dq
(1-\cos\vartheta_{pq})\left(\frac{1-|z(q,t)|^2}{1+|z(q,t)|^2} \right)
\nonumber
\end{equation}
and 
\begin{equation}
\beta(p,t)=\frac{1}{2} \frac{\delta m^2}{2p}\sin{2\theta} +
\sqrt{2}G_F \int dq
(1-\cos\vartheta_{pq}) 
\left(\frac{z(q,t)}{1+|z(q,t)|^2} \right) . \nonumber 
\end{equation}
Defining
\begin{equation}
z(p,t)=\frac{\psi_x(p,t)}{\psi_e(p,t)},
\nonumber
\end{equation}
with the auxiliary condition $ \>\>\> |\psi_e|^2+|\psi_x|^2=1$, Eq. (\ref{zeq}) reduces to  

\begin{equation}
i\frac{\partial}{\partial t} \left( \begin{array}{c} \psi_e
\\ \psi_x \\ \end{array} \right) =
\frac{1}{2}\left(%
\begin{array}{cc}
  A+D-\Delta\cos{2\theta} & D_{e\mu}+\Delta\sin{2\theta} \\
  D_{\mu e}+\Delta\sin{2\theta} & -A-D+\Delta\cos{2\theta} \\
\end{array} \right)
\left( \begin{array}{c} \psi_e \\
\psi_x \\ \end{array} \right) \label{eveq}
\end{equation}
where 
\begin{equation}
\Delta=\frac{\delta m^2}{2p}, \ \ \ \ \ \ \ A=\sqrt{2} G_F N_e, \nonumber
\end{equation}
\begin{equation}
D=\sqrt{2}G_F \int dq
(1-\cos\vartheta_{pq})\left[\left(|\psi_e(q,t)|^2-|\psi_x(q,t)|^2\right) 
\right] ,\nonumber
\end{equation}
\begin{equation}
D_{ex}=2\sqrt{2}G_F \int dq
(1-\cos\vartheta_{pq})\left(\psi_e(q,t)\psi^*_x(q,t) \right) .  \nonumber
\end{equation}

In the stationary point approximation to the full quantum mechanical problem, the test neutrino interacts with an "average field" representing the effect of all the other neutrinos. This approximation is analogous to the random phase approximation (RPA), widely used in many-body physics. In the RPA one can approximate product of two commuting arbitrary operators $\hat{\cal O}_1$ and $\hat{\cal O}_2$ as
\begin{equation}
\label{rpa}
\hat{\cal O}_1 \hat{\cal O}_2 \sim \hat{\cal O}_1 \langle \xi |
\hat{\cal O}_2 | \xi \rangle + \langle \xi | \hat{\cal O}_1 |
\xi \rangle \hat{\cal O}_2 -
\langle \xi | \hat{\cal O}_1 |
\xi \rangle
\langle \xi | \hat{\cal O}_2 |
\xi \rangle , \nonumber
\end{equation}
provided that 
\begin{equation}
\label{rpacon}
\langle \xi | \hat{\cal O}_1  \hat{\cal O}_2 | \xi \rangle =
\langle \xi | \hat{\cal O}_1 |
\xi \rangle
\langle \xi | \hat{\cal O}_2 |
\xi \rangle.
\end{equation}
This approximation reduces $H_{\nu\nu}$ to a {\it one-body} Hamiltonian:
\begin{equation}
\label{apprpa}
{\mathcal H}_{\nu \nu} \sim 2\frac{\sqrt{2}G_F}{V} \int d^3p \> d^3q
\> R_{pq} \> \left( J_0(p) \langle J_0(q) \rangle 
+  \frac{1}{2}
J_+(p) \langle J_-(q) \rangle + \frac{1}{2} J_-(p) \langle J_+(q)
\rangle\right).  \nonumber
\end{equation} 

The pre-exponential determinant obtained in the stationary phase approximation to the path integral is rather complicated and an explicit evaluation is not yet available in the literature. For simplicity in the discussion above we omitted antineutrinos. Antineutrinos can be included by introducing a second set of SU(2) algebras \cite{Balantekin:2006tg}. Similarly incorporating three flavors requires introduction of SU(3) algebras \cite{Sawyer:2005jk}. Both extensions are straightforward, but tedious. 

Introducing the polarization vectors
\begin{equation}
\label{z1}
 P_i(q) = {\rm Tr} ( J_i(q) \rho) 
\end{equation}
with the density matrix 
\begin{equation}
\rho =  \left( \begin{array}{cc}
\rho_{ee} & \rho_{ex} \\ \rho_{xe} & \rho_{xx} \end{array} \right)
=\frac{1}{2} \left( P_0 + {\bf P} \cdot {\bf \sigma} \right) , \nonumber 
\end{equation}
one can show that, in RPA and including antineutrinos,  the evolution equation (\ref{eveq}) takes the commonly-used forms 
\begin{equation} 
\label{rpa1}
\partial_r{\mathbf P}_p = \left\{
+{\frac{\delta m^2}{2p}(\sin 2\theta
{\mathbf {\hat x}} - \cos 2 \theta {\mathbf {\hat z}}) } + \sqrt2\,G_{\rm F} \left[ N_{e} \hat{\bf z}
+ \int d{\bf q} \left( 1-\cos\vartheta_{pq} 
\right)
({\mathbf P_{\bf q}}- {\mathbf{\overline P}}_{\bf q}) \right]
\right\}
\times
{\mathbf P}_p \nonumber
\end{equation}
and 
\begin{equation}
\label{rpa2} 
\nonumber \partial_r{\mathbf{\overline P}}_p = \left\{
-{\frac{\delta m^2}{2p}(\sin 2\theta
{\mathbf {\hat x}} - \cos 2 \theta {\mathbf {\hat z}}) } + \sqrt2\,G_{\rm F} \left[ N_{e} \hat{\bf z}
+  \int d{\bf q} \left( 1-\cos\vartheta_{pq} 
\right) ({\mathbf P_{\bf }}_{\bf q}- {\mathbf{\overline P}}_{\bf q}) \right]
\right\}
\times {\mathbf{\overline P}}_p . \nonumber 
\end{equation}

\section{Invariants of the neutrino-neutrino interaction Hamiltonian}
 
We consider the limit where the neutrino-neutrino interactions dominate and neglect the interactions with the background electrons and positrons.  The Hamiltonian then becomes 
\[
\hat{H} = 
\sum_p\frac{\delta m^2}{2p}\hat{B}\cdot\vec{J}_p
+\frac{\sqrt{2}G_{F}}{V}\sum_{\mathbf{p},\mathbf{q}}\left(1-\cos\vartheta_{\mathbf{p}\mathbf{q}}\right)\vec{J}_{\mathbf{p}}\cdot\vec{J}_{\mathbf{q}} .
\]
We further limit this discussion to the so-called single-angle approximation where all the $\vartheta$ are the same.  Defining $\mu= (1- \cos \vartheta) \frac{\sqrt{2}G_{F}}{V}$, $\tau=\mu t$, and    $ \omega_{p}=\frac{1}{\mu}\frac{\delta m^{2}}{2p}$, the Hamiltonian takes the form 
\begin{equation}
\label{ham}
\hat{H} = \sum_p\omega_{p}\hat{B}\cdot\vec{J}_p
+\vec{J}\cdot\vec{J} . 
\end{equation}

This Hamiltonian of Eq. (\ref{ham}) preserves the \emph{length of each spin}
\[
\hat{L}_p=\vec{J}_{p}\cdot\vec{J}_{p}~,
\qquad\qquad \left[ \hat{H}, \hat{L}_p\right]=0~,
\]
as well as the \emph{total spin component} in the direction of the "external magnetic field", $\hat{B}$  
\begin{equation}
\label{c0}
\hat{C}_0 =\hat{B}\cdot\vec{J}~, \qquad\qquad\quad \left[\hat{H},\hat{C}_0\right]=0~ .
\end{equation}
It is possible to show that \cite{Pehlivan:2011hp} the collective neutrino Hamiltonian of Eq. (\ref{ham}) has the following constants of motion: 
\begin{equation}
\label{invar}
\hat{h}_{p} = \hat{B}\cdot\vec{J}_p+2\sum_{q\left(\neq p\right)}\frac{\vec{J}_{p}\cdot\vec{J}_{q}}{\omega_{p}-\omega_{q}}. 
\end{equation}
The individual neutrino spin-length discussed before in an independent invariant. However 
$
\hat{C}_0=\sum_{p}\hat{h}_{p}$. 
The Hamiltonian itself is also a linear combination of these invariants. 
\[
\hat{H}=\sum_{p}w_{p}\hat{h}_{p}+\sum_{p} \hat{L}_{p}~.
\]

The maximal value of the neutrino flavor isospin quantum number is $J_{\mbox{\tiny max}}=N/2$, where $N$ is total number of neutrinos. For example a state with all electron neutrinos is
$
|\nu_e \; \nu_e \; \nu_e \; \dots \rangle = | J_{\mbox{\tiny max}} \; \; J_{\mbox{\tiny max}}\rangle_f $. It is also easy to show that matter and flavor bases are connected with the unitary transformation of Eq. (\ref{mixalg}): 
$
| J_{\mbox{\tiny max}} \; \; J_{\mbox{\tiny max}}\rangle_f = \hat{U}^\dagger| J_{\mbox{\tiny max}} \; \; J_{\mbox{\tiny max}}\rangle_m
$. One has $
|J_{\mbox{\tiny max}}\;\;J_{\mbox{\tiny max}}\rangle_m= \prod_{\mathbf{p},s}a_{1}^{\dagger}(\mathbf{p},s)\;|0\rangle$ and 
$|J_{\mbox{\tiny max}}\;-J_{\mbox{\tiny max}}\rangle_m= \prod_{\mathbf{p},s}a_{2}^{\dagger}(\mathbf{p},s)\;|0\rangle 
$, corresponding to the energies 
$
E_{(+J_{\mbox{\tiny max}})}=-\sum_{p}\frac{n_p}{2}\omega_{p}+J_{\mbox{\tiny max}}\left(J_{\mbox{\tiny max}}+1\right)
$
and 
$
E_{(-J_{\mbox{\tiny max}})}=\sum_{p}\frac{n_p}{2}\omega_{p}+J_{\mbox{\tiny max}}\left(J_{\mbox{\tiny max}}+1\right) ,
$
respectively. Energy eigenvalues can be obtained by introducing the raising operator \cite{Pehlivan:2011hp}
\[
\mathcal{Q}^\pm(\xi) = 
= \sum_p \frac{1}{\omega_p-\xi} \left(\cos^2 \theta \hat{J}_p^\pm +\sin2\theta \hat{J}_p^0 -\sin^2 \theta \hat{J}_p^\mp\right) . 
\]
Applying the operator to the state $|J \; -J \rangle_m$ yields a term proportional to $\mathcal{Q}^+(\xi)|J \; -J \rangle_m$ and an additional term. Setting the coefficient of this latter term to zero gives the Bethe ansatz equation $\sum_p\frac{-j_p}{w_p-\xi}=-\frac{1}{2}$ that needs to be satisfied if $\mathcal{Q}^+(\xi)|J \; -J \rangle_m$ is an eigenstate. The most general eigenstate is 
\[
|\xi_1,\xi_2,\dots\xi_\kappa\rangle\equiv \mathcal{Q}^+(\xi_1)\mathcal{Q}^+(\xi_2)\dots\mathcal{Q}^+(\xi_\kappa)|J \; -J \rangle_m
\]
which corresponds to the eigenvalue 
\[
E(\xi_1, \xi_2,\dots,\xi_\kappa)=E_{(-J)}-\sum_{\alpha=1}^\kappa \xi_{\alpha}-\kappa(2J-\kappa+1)~,
\]
provided that the following Bethe ansatz equations are satisfied: 
\[
\sum_{p}\frac{-j_p}{\omega_p-\xi_\alpha}=-\frac{1}{2}+\sum_{\substack{\beta=1\\ \left(\beta\neq\alpha\right)}}^{\kappa}\frac{1}{\xi_{\alpha}-\xi_{\beta}}~.
\]

Using the polarization vector, $\vec{P}_{\mathbf{p},s}=2\langle\vec{J}_{\mathbf{p},s}\rangle$, of Eq. (\ref{z1}) one can write the Hamiltonian of Eq. (\ref{ham}) as 
\begin{equation}
\label{rpasham}
\hat{H}\sim\hat{H}^{\mbox{\tiny RPA}} =  \sum_p\omega_{p}\hat{B}\cdot\vec{J}_p
+\vec{P}\cdot\vec{J} .
\end{equation}
This Hamiltonian yields the Heisenberg equations of motion for the operators $\vec{J}_{p}$: 
\begin{equation}
\label{eqmot}
\frac{d}{d\tau}\vec{J}_{p} = -i[\vec{J}_p,\hat{H}^{\mbox{\tiny RPA}}]=(\omega_p\hat{B}+\vec{P})\times\vec{J}_{p}.
\end{equation}
Applying the RPA consistency conditions of Eq. (\ref{rpacon}) to Eq. (\ref{eqmot}) one obtains the equations of motion in the RPA: 
\[
\frac{d}{d\tau}\vec{P}_{p} = (\omega_{p}\hat{B}+\vec{P})\times\vec{P}_{p} . 
\]
Note that this is the single-angle limit of Eq. (\ref{rpa1}) only with neutrinos. It turns out that expectation value of the invariants, Eq. (\ref{invar}), of the exact many-body Hamiltonian
\begin{equation}
\label{ip}
I_p=2\langle \hat{h}_p \rangle  = \hat{B}\cdot\vec{P}_p+\sum_{q\left(\neq p\right)}\frac{\vec{P}_{p}\cdot\vec{P}_{q}}{\omega_{p}-\omega_{q}},
\end{equation}
is an invariant of the RPA Hamiltonian: 
\[
\frac{d}{d\tau} I_p =0 .
\]
Introduction of antineutrinos is again straightforward utilizing a second SU(2) algebra (denoted below with a tilde over the appropriate quantity).  The conserved quantities for the single-angle Hamiltonian with both neutrinos and antineutrinos 
\[
H = \sum_{p}\omega_p\hat{B}\cdot\vec{J}_p
+\sum_{\bar{p}}\omega_{\bar{p}}\hat{B}\cdot\vec{\tilde{J}}_p+\left(\vec{J}+\vec{\tilde{J}}\right)\cdot\left(\vec{J}+\vec{\tilde{J}}\right) 
\]
are 
\[
\hat{h}_{p} = \hat{B}\cdot\vec{J}_p+2\sum_{q\left(\neq p\right)}\frac{\vec{J}_{p}\cdot\vec{J}_{q}}{\omega_{p}-\omega_{q}}+2\sum_{\bar{q}}\frac{\vec{J}_{p}\cdot\vec{\tilde{J}}_{\bar{q}}}{\omega_{p}-\omega_{\bar{q}}}
\]
for each neutrino energy mode $p$ and 
\[
\hat{h}_{\bar{p}} =\hat{B}\cdot\vec{\tilde{J}}_p+2\sum_{\bar{q}\left(\neq\bar{p}\right)}\frac{\vec{\tilde{J}}_{\bar{p}}\cdot\vec{\tilde{J}}_{\bar{q}}}{\omega_{\bar{p}}-\omega_{\bar{q}}}+2\sum_{q}\frac{\vec{\tilde{J}}_{\bar{p}}\cdot\vec{J}_{q}}{\omega_{\bar{p}}-\omega_{q}}~.
\]
or each antineutrino energy mode. In the RPA these take the form 
\[I_p=2\langle\hat{h}_{p}\rangle =\hat{B}\cdot\vec{P}_p+\sum_{q\left(\neq p\right)}\frac{\vec{P}_{p}\cdot\vec{P}_{q}}{\omega_{p}-\omega_{q}}+\sum_{\bar{q}}\frac{\vec{P}_{p}\cdot\vec{\tilde{P}}_{\bar{q}}}{\omega_{p}-\omega_{\bar{q}}}
\]
and
\[
I_{\bar{p}}=2\langle\hat{h}_{\bar{p}}\rangle = \hat{B}\cdot\vec{\tilde{P}}_{\bar{p}}+\sum_{\bar{q}\left(\neq\bar{p}\right)}\frac{\vec{\tilde{P}}_{\bar{p}}\cdot\vec{\tilde{P}}_{\bar{q}}}{\omega_{\bar{p}}-\omega_{\bar{q}}}+\sum_{q}\frac{\vec{\tilde{P}}_{\bar{p}}\cdot\vec{P}_{q}}{\omega_{\bar{p}}-\omega_{q}} , 
\]
respectively. 

Recently a lot of attention was paid to the spectral splitting (or spectral swapping) phenomenon 
\cite{Raffelt:2007cb,Dasgupta:2010cd,Duan:2007bt,Fogli:2009rd,Galais:2011gh}. To explore the origin of this phenomenon we note that 
the expectation value of the invariant in Eq. (\ref{c0}) which can be written in terms of the expectation value of the invariants in Eq. (\ref{ip}), 
\begin{equation}
\label{c0ex}
\sum_p I_p = \mathbf{B} \cdot \mathbf{P}, 
\end{equation}
is not conserved by the RPA Hamiltonian, Eq. (\ref{rpasham}). Its conservation needs to be enforced using a Lagrange multiplier. Since $\sum_pI_p$ is proportional to $\hat{J}^0$, one needs to diagonalize the quantity
\begin{eqnarray}
\label{lm}
\hat{H}^{\mbox{\tiny RPA}}+\omega_c\hat{J}^0 &=& \sum_{p}(\omega_c-\omega_p)\hat{J}_p^0+\vec{\mathcal{P}}\cdot\vec{J} \nonumber \\
&=& \sum_{\mathbf{p},s}  
2\lambda_p  \hat{U}^{\prime\dagger} \hat{J}^0_p \hat{U}^{\prime},  \nonumber
\end{eqnarray}
where the transforming operator is parameterized as 
\begin{equation}
\label{digtr}
\hat{U}^\prime=e^{\sum_p z_p J_p^+}\;e^{\sum_p \ln(1+|z_p|^2) J_p^0}\;e^{-\sum_p z_p^* J_p^-} 
\end{equation}
with 
\[
z_p=e^{i\delta}\tan{\theta_p}
\]
and 
\[
\cos{\theta_p} =\sqrt{\frac{1}{2}\left(1+\frac{\omega_c-\omega_p+\mathcal{P}^0}{2\lambda_p}\right)}. 
\]
This operator transforms matter-basis creation and annihilation operators into {\it quasi-particle} creation and annihilation operators:    
\begin{eqnarray}
\alpha_1(\mathbf{p},s)&=&\hat{U}^{\prime\dagger} a_1(\mathbf{p},s)\hat{U}^{\prime} 
= \cos{\theta_p} \; a_1(\mathbf{p},s) -e^{i\delta}\sin{\theta_p} \; a_2(\mathbf{p},s) \nonumber \\
\alpha_2(\mathbf{p},s)&=&\hat{U}^{\prime\dagger} a_2(\mathbf{p},s)\hat{U}^{\prime} 
=  e^{-i\delta}\sin{\theta_p} \; a_1(\mathbf{p},s)+\cos{\theta_p} \; a_2(\mathbf{p},s)\nonumber
\end{eqnarray}
so that we obtain a diagonal Hamiltonian: 
\[
\hat{H}^{\mbox{\tiny RPA}}+\omega_c\hat{J}^0= \sum_{\mathbf{p},s}  
\lambda_p \left( \alpha_1^\dagger(\mathbf{p},s)\alpha_1(\mathbf{p},s)-\alpha_2^\dagger(\mathbf{p},s)\alpha_2(\mathbf{p},s)   \right) . 
\]

Let us assume that initially ($\lim \mu \rightarrow \infty$) there are more $\nu_e$'s and all neutrinos are in flavor eigenstates. We then have 
\[
\lim \cos\theta_p = \lim \sqrt{\frac{1}{2}\left(1+\frac{P^0}{|\vec{P}|}\cos{2\theta}\right)} = \cos \theta,
\]
i.e., the diagonalizing transformation of Eq. (\ref{digtr}) reduces into the neutrino mixing transformation of Eq. (\ref{mixalg}) and the total Hamiltonian of Eq. (\ref{lm}) is diagonalized by the flavor eigenstates:  
\[
\alpha_1(\mathbf{p},s)=\hat{U}^\dagger a_1(\mathbf{p},s)\hat{U} = a_e(\mathbf{p},s) .
\]
After neutrinos propagate to a region with very low neutrino density ($\mu \rightarrow 0$) one gets  
\[
\cos\theta_p=\sqrt{\frac{1}{2}\left(1+\frac{\omega_c-\omega_p}{|\omega_c-\omega_p|}\right)} 
\Rightarrow \left\{ \begin{array}{rl} 
1 &  \omega_p < \omega_c\\
 0 & \omega_p > \omega_c
\end{array} \right.
\]
yielding
\[
\alpha_1(\mathbf{p},s)=\hat{U}^\dagger a_1(\mathbf{p},s)\hat{U} \Rightarrow \left\{ \begin{array}{rl} 
a_1(\mathbf{p},s)  &  \omega_p < \omega_c\\
- a_2(\mathbf{p},s)  & \omega_p > \omega_c
\end{array} \right. ,
\]
i.e. neutrinos with $\omega_p < \omega_c$ and $\omega_p > \omega_c$ evolve into different mass eigenstates. In Ref. \cite{Raffelt:2007cb} it was shown that such an evolution leads to spectral splits. 

\section{ Concluding Remarks}

Neutrino propagation through the dense media in the core-collapse supernovae probes many interesting collective effects.  Because of the neutrino-neutrino interactions, this many-body system is intrinsically non-linear, it can be linearized only in certain cases \cite{Banerjee:2011fj}. 
We examined this many-neutrino system  both from the exact many-body perspective and from the point of view of an effective one-body description formulated with the application of the RPA method. To achieve this goal we exploited mathematical similarities between the neutrino-neutrino interaction Hamiltonian and the BCS pairing Hamiltonian. (Indeed, the N-mode collective oscillations of the neutrinos are related to the $m$-spin solutions of the BCS model \cite{yuz}).  
In the limit of the single angle approximation, both the many-body and the RPA pictures possess many constants of motion manifesting the existence of associated dynamical symmetries in the system. 
Judicious use of these invariants could certainly help numerical calculations \cite{Duan:2008eb}. 

We treated the two-flavor neutrino-neutrino interaction in the single-angle approximation, and provided an interpretation of the critical energy in the spectral swap/split phenomenon as the Lagrange multiplier of the number conservation constraint. 
Recent numerical work with three flavors in the multi-angle approximation uncovers significant differences between single- and multi-angle formulations \cite{Cherry:2010yc}. In particular, multi-angle formulation is found to reduce the adiabaticity of flavor evolution in the normal neutrino mass hierarchy, resulting in lower swap energies. Thus it seems that single-angle approximation seems to be sufficient in some cases, but inadequate in other situations.  

Other questions remain regarding the many-body behavior of the neutrino system. For example, in the calculations so far neutrinos are assumed to be emitted half-isotropically (only outward-moving modes are occupied with backward-moving modes being empty). However, recent realistic supernova simulations suggest that neutrino angular distributions are not half-isotropic \cite{Ott:2008jb}. Flavor-dependent angular distributions may lead to multi-angle instabilities 
\cite{Sawyer:2008zs,Mirizzi:2011tu}.  Future work could uncover even more interesting features of this many-neutrino system. 

\section*{Acknowledgments}
This work was supported in part
by the U.S. National Science Foundation Grant No. PHY-0855082, and 
in part by the University of Wisconsin Research Committee with funds
granted by the Wisconsin Alumni Research Foundation. I would like to thank the organizers of HANSE 2011 for their hospitality.

 

\begin{footnotesize}



%

\end{footnotesize}


\end{document}